\documentclass{osa-article}

\journal{ome}

\articletype{Research Article}

\usepackage{amsmath}

\begin{document}

\title{Non-spectroscopic sensing enabled by electro-optical reservoir computer}

\author{Gleb Anufriev,\authormark{1} David Furniss,\authormark{1} Mark Farries,\authormark{2} and Sendy Phang\authormark{1,*}}

\address{\authormark{1}George Green Institute for Electromagnetics Research, University of Nottingham, NG7 2RD, UK \\
\authormark{2} Gooch \& Housego, Broomhill Way, Torquay, TQ2 2QL, UK }
\email{\authormark{*}sendy.phang@nottingham.ac.uk} 



\begin{abstract}
A fully bio-inspired apparatus to perform non-spectroscopic sensing to discriminate, and determine the constituent concentrations of a chemical mixture is proposed. Here, fully bio-inspired means that it is comprised of a biomimetic sensor and a neuromorphic signal processor. The sensor is similar to the human eye in terms of its trichromat architecture and overlapping spectral response. The information processor is a neuromorphic system based on an electro-optical implementation of a reservoir computer. A time-stepping signal algorithm based on the $\mathcal{Z}$-bilinear transformation has been developed to realistically simulate the electro-optical reservoir computer; this demonstrates the discrimination and chemical concentration determination tasks. We believe such an apparatus offers potential benefits in areas in which chemical composition needs to be monitored in real time, for example in chemical processing, and food-beverage (fermentation/brewing) and environmental monitoring.
\end{abstract}

\section{Introduction}
The human nervous system is an efficient and powerful ‘biological computer’. It consumes 20 Watts of power, $<1\%$ of the power consumption of the average kettle and can process an enormous stream of information in real-time \cite{levy2020computation,wilson2018your,BeaulieuLaroche2018}. Key to the vast capability of the human nervous system lies is how and where such information is processed. An electronic computer processes information at a centralised processing unit (von Neumann architecture) with a separate memory device. The human nervous system is, however, based on a distributed processing architecture in which each neuron “remembers and processes” a stream of (electro-chemical) signals as it travels to the brain for “inferring”. Inspired by the distributed computing architecture of the human nervous system, several neuromorphic computing architectures have been developed\cite{Prucnal2017,shastri2017principles,de2017progress,Tanaka2019}. The neuromorphic computing system aims to realise the distributed computing architecture of the human nervous system as hardware, enabling a hardware-level integration of Artificial Intelligence. 

Photonic Reservoir Computer (PhRC) is a relatively new member of the neuromorphic computing family implemented as a physical photonic/optical system.  {Implementing a reservoir computer in photonics leads to a design for very high speed processing exploiting the availability of ultrawide bandwidth optical communication commponents.} It offers a fundamentally different approach to other neuromorphic photonic systems, in terms of \textbf{both} architecture and how the system is trained (or optimised) \cite{Prucnal2017,shastri2017principles,de2017progress,Tanaka2019,christensen20212021}. PhRC \textbf{only} trains/optimises the Read-Out whilst retaining the characteristic of the kernel as a random and semi-chaotic system, whereas other neuromorphic systems require both kernel and read-out training/optimisation\cite{Prucnal2017,shastri2017principles,de2017progress,Tanaka2019}. Such a training approach makes the PhRC more robust and less sensitive to perturbation. Consequently, the PhRC offers a more straightforward implementation and scalability than other neuromorphic photonic architectures\cite{Prucnal2017,shastri2017principles,de2017progress,Tanaka2019,christensen20212021}. Several implementations of PhRC have been reported, for example as a swirl configuration of semiconductor amplifiers \cite{vandoorne2011parallel,Vandoorne2008}, or as a photonic crystal cavity \cite{Laporte2018,Phang2020,phangiet2020}. In this paper, we consider an electro-optical reservoir computer (EORC) implementation with a single delay line and a Mach-Zehnder modulator (MZM) as the non-linear neuron activation function \cite{appeltant2011information,Paquot2012}. 

Humans’ powerful computing capacity has mainly been used to interpret sensory information, $~99.99\%$ of the information processing capacity is for sensory processing while $<0.01\%$ is for attention, decision-making, motion and language \cite{Wu2016}. Motivated by such a fact, in our previous work\cite{Phang2020,phangiet2020,Phang2021}, we have numerically demonstrated a new methodology to process ultrafast high-throughput molecular vibrational signals (spectroscopic) using a chaotic photonic crystal billiard system as its kernel. The present paper reports an alternative method which is fully-bio-inspired, meaning that the sensor system is biomimetic of the human eyes and its output will be processed through a neuromorphic EORC system. Numerical simulation successfully demonstrates that both chemical discrimination and concentration determination tasks can be achieved. 

The paper is organised as follows: Section 2 describes the system as a whole and its composition, namely the biomimetic sensor and the EORC and the modelling approach used. Section 3 discusses the results obtained. The first part considers the dynamics of the EORC system, which is modelled using a realistic model considering the filtering effects observed in a practical setup. The second part demonstrates the application of such a system to perform the chemical discrimination and concentration determination tasks. The impact of EORC parameters on the accuracy is investigated and discussed.  Some concluding remarks are then provided in Section 4.   

\section{Biomimetic sensing by neuromorphic signal processing}
\begin{figure}[tbp]
    \centering
    \includegraphics[width=0.97\textwidth]{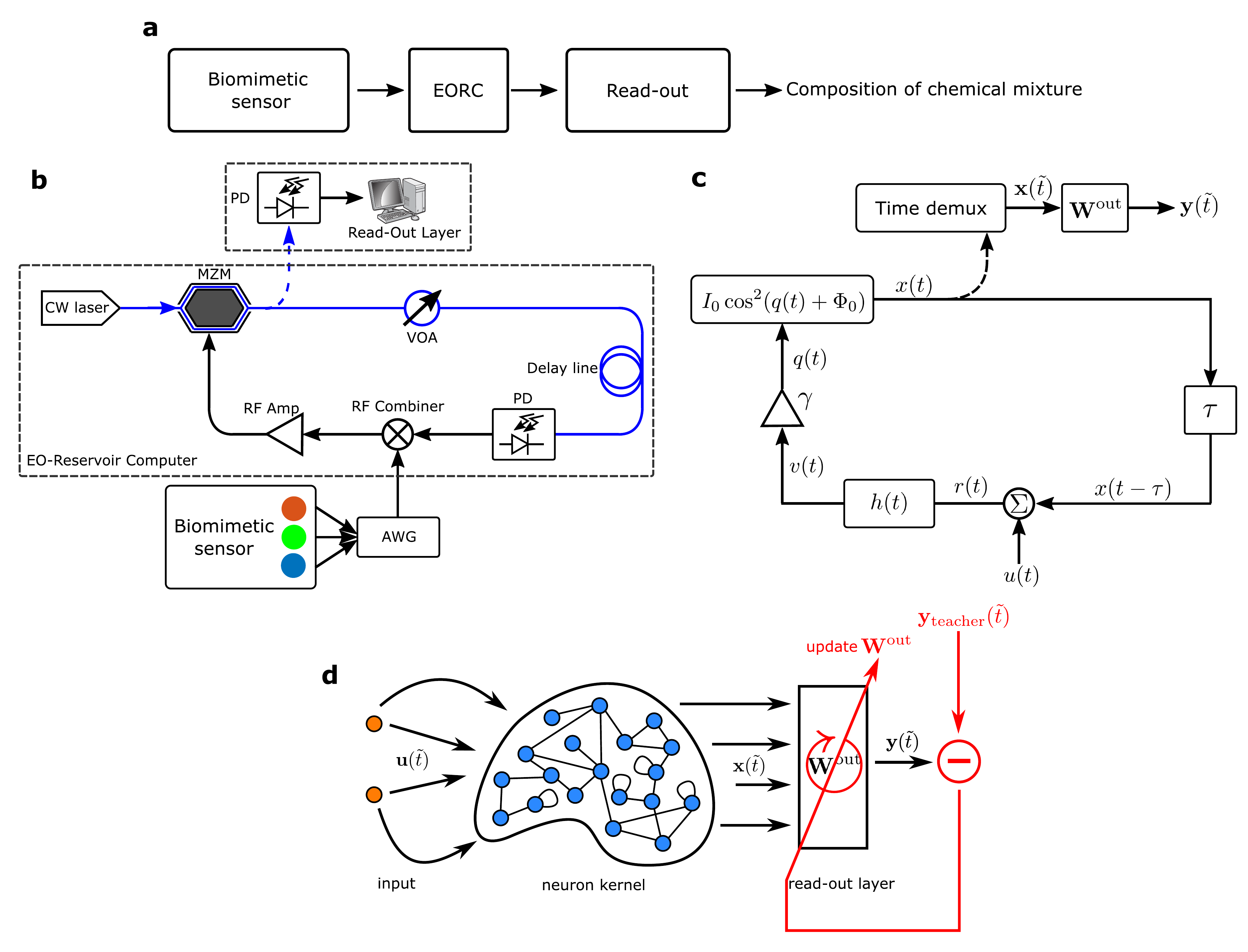}
    \caption{Schematic illustration of the system as a whole; (b) Implementation of (a) in a single physical node configuration with electro-optic with Mach-Zehnder modulator (MZM) as nonlinear neuron activation kernel; The multi-channels output of the biomimetic sensor has been representatively shown in a Red-Green-Blue (RGB) colouring scheme. (c) Signal flow diagram translating the flow of signal in (b); (d) Schematic of the architecture of algorithmic reservoir computer.}
    \label{fig:f1}
\end{figure}

Figure \ref{fig:f1}(a) schematically shows the apparatus considered in this paper. It is comprised of a biomimetic sensor serving as the sensing component, an electro-optic reservoir computer (EORC) serving as signal processing component and a read-out component serving as an inferring component of the EORC output.  Figure \ref{fig:f1}(b) schematically depicts the implementation of Fig.~\ref{fig:f1}(a) using a single delay-line configuration and a Mach-Zehnder modulator (MZM) as the nonlinear neuron activation kernel. The biomimetic sensor supplies a three-channel reading to the arbitrary waveform generator which performs time-multiplexing on the sensing information for input to the EORC system. 

\subsection{Biomimetic sensor system}
This sub-section describes the operating principle and the modelling procedures of the sensing component used in this work. It is inspired by the biomimetic design which previously has been demonstrated successfully for chemical \cite{Major2014,Major2016,Major2018} and biological sample \cite{Ewing2020} detection. The biomimetic sensor mimics the operation of the human eyes,  which are capable of discriminating over 100,000 shades of (visible) colour over an environmental background despite being equipped with only three main types of cone cells \cite{Flinkman2012,Linhares2008}.  {Key to this ability is the spectral sensitivity of these cone cell types which are sprectrally broad and overlapping \cite{Schnapf1987}. Research on colour sensitivity of mantis shrimp (\textit{Haptosquilla trispinosa}), which has 12 different types of photosensitive cell covering different light wavelengths and polarisations, however, found that higher number of unique photoreceptors does not mean an improvement in colour discimination \cite{Thoen2014}; Mantis's spectral discrimination $\Delta \lambda$ is in the region of 15 to 25 nm in comparison to humans' which is in the region of 1 to 8 nm \cite{Thoen2014,DeValois1968} over the visible light region.} In \cite{Major2014,Major2016,Major2018,Ewing2020}, a trichromat, three-colour, biomimetic sensor has been constructed in the mid-infrared range of light and a multi-variate statistics tool based on a patented comparative-discrimination spectral detection (CDSD) technique \cite{Major2018,poutous2018comparative} has been used to perform the vectorial signal processing. In the present work, we have adopted a biomimetic sensor with a similar trichromat architecture, but the EORC is used for signal processing. The EORC will be described in sub-section 2.2.

\begin{figure}[tbp]
    \centering\includegraphics[width=0.85\textwidth]{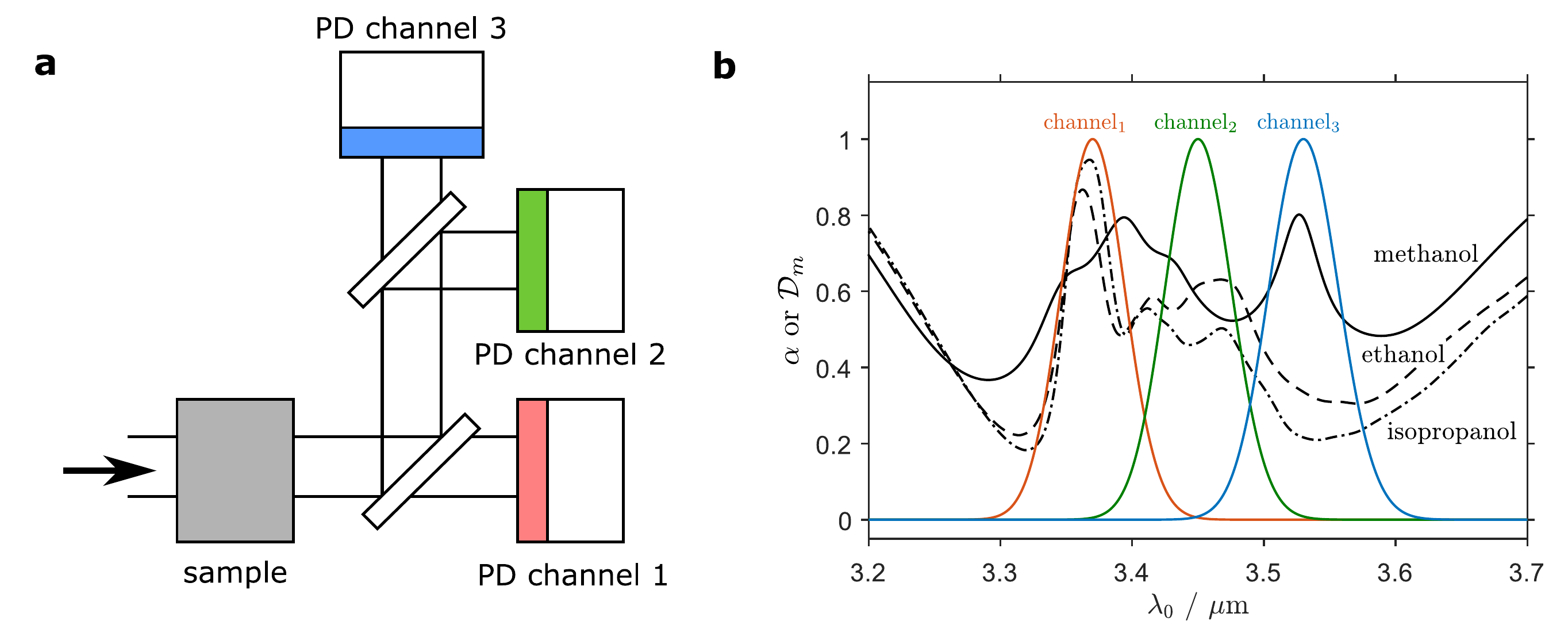}
    \caption{Schematic design of the three-channel biomimetic sensor; (b) the absorption spectra $\alpha$ and the filter response $\mathcal{D}_m(f)$ of the photodetector (PD).}
    \label{fig:f2}
\end{figure}

As mentioned, the biomimetic sensor used in this paper comprised of three channels whose spectral responses are different, see Fig. \ref{fig:f2}. Figure \ref{fig:f2}(a) schematically shows the implementation of the biomimetic sensor constructed in\cite{Major2014,Major2016,Major2018}. The photodetectors are equipped with a filter at different central wavelength, in Fig. \ref{fig:f2}(a) the filters have been representatively shown in Red-Green-Blue (RGB) layer in front of the detectors. In the present work, we have specifically considered that each channel has a Gaussian spectral response peaking at different wavelengths, i.e., 3.37, 3.45 and 3.53 $\mu$m and with a half-power bandwidth (HPBW) of 60 nm. Such sensors are readily available, for example as  {dielectric} filtered photodetectors\cite{infratec}. The specific parameter values of the sensor have been chosen as such they are broad and spectrally overlapped\cite{Major2014,Major2016,Major2018,Ewing2020}, see Fig.~\ref{fig:f2}(b). The choice of peak spectral response around the 3 µm of the mid-infrared range is motivated by the richness of carbon-hydrogen stretching fingerprints in this region\cite{Borondics2018}. Note that in previous works\cite{Major2014,Major2016,Major2018,Ewing2020}, the sensor has also operated at similar wavelengths of interest.   

For demonstration, in this work methanol, ethanol and isopropanol and their mixture will be considered as chemicals to be detected. The  {complex-valued} frequency-dependent refractive indices of the pure substances have been taken from \cite{myers2018accurate} whereas the refractive indices of mixture are calculated using the Lorentz-Lorenz rule \cite{lorentz1916theory,An2015,Mayerhfer2020} by, 
\begin{align}
    \frac{n^2-1}{n^2+2} = \sum_{k} \varrho_k \frac{n_k^2-1}{n_k^2+2}
    \label{eq1}
\end{align}
in (\ref{eq1}), $n$ and $n_k$ are the refractive indices of the mixture and the pure components, respectively. The $\varrho_k$ denotes the relative volume mixture of component $k$, with $\sum_k\varrho_k =1$. The transmitted power ratio, $T$, upon absorption by the sample, is calculated in the absence of any scattering and loss due to the sample holder as \begin{align}
    T(f,n) = \exp \left(-\frac{4\pi f |n_I(f)| \ell}{c_0} \right)
    \label{eq2}
\end{align}
In (\ref{eq2}), $f$ and $c_0$ are the frequency and speed of light respectively. Note that the refractive index $n=n_R+i|n_I|$ convention is used. The thickness of the sample $\ell=5$ $\mu$m is fixed throughout this work. It is important to note that the treatment used here to calculate the transmitted light in the mixture sample is consistent with the Maxwell’s equations of electromagnetics; for a detailed discussion readers are referred to Review \cite{Mayerhfer2020}. For reference, the absorption, $\alpha=1-T$, of pure methanol, ethanol and isopropanol is overlaid over the spectral response of the biomimetic sensors in Fig. \ref{fig:f2}(b). The power observed by the photodetector is calculated by,  
\begin{align}
    P_m(n)=\int_0^\infty \mathcal{S}_{\mathrm{in}}(f) \mathcal{D}_m(f) T(f,n) \mathrm{d}f, \qquad m=1,2,3
    \label{eq3}
\end{align}
where, $\mathcal{D}_m(f)$ defines the spectral response of the photodetector at channel $m$ of the sensor. For demonstration purposes, the sensor is assumed to be equipped with a broadband light source with a top-hat spectral power density $ \mathcal{S}_{\mathrm{in}}$ profile from 3.1 to 3.7 $\mu$m in wavelength that is normalised, i.e., $\int \mathcal{S}_{\mathrm{in}}(f) \mathrm{d}f = 1$ W.  {For practical realisations, photo-luminescent emission from rare earth doped optical fibre \cite{Sojka2019} or a supercontinuum light-source \cite{petersen2014mid,Sylvestre2021} can be used.} Figure~\ref{fig:f9}(a) shows the pseudo-colouring of the biomimetic sensor output $\mathbf{P}(n)=[P_1;P_2;P_3]^\mathrm{T}$ following the Red-Green-Blue (RGB) colour scheme. To allow better contrast for visualisation, the sensor output has been normalised during the pseudo-colouring, that is $\mathbf{P}_\mathrm{visual}(n)= \mathbf{P}(n)/\max[\{\mathbf{P}\}]$, where $\{\mathbf{P}\}$ denotes the whole set of $\mathbf{P}(n)$. Note that only on Figure \ref{fig:f9}(a), normalisation has been performed. Figure \ref{fig:f9} will be discussed in detail in Section 3.3.

\subsection{Electro-optical reservoir computer (EORC)}
In the previous sub-section, the operating principle and model description of the biomimetic sensor was described. In this sub-Section, its physical implementation using electro-optical (EO) modulator system and its numerical model are described. The EORC is a physical implementation of the Reservoir Computer (RC) computational framework which is a generalisation of a traditional artificial neural-network (ANN) based on the recurrent neural-network (RNN) \cite{Tanaka2019,ganesh2020rebooting,VanderSande2017,Lukoeviius2009}. For detail on the framework of the RC, the reader is referred to \cite{Lukoeviius2009}. Figure \ref{fig:f1}(d) shows a schematic of the architecture of an RC scheme. An RC system consists of three main parts, namely an input, a kernel and a Read-Out layer. The origin of the name `Reservoir Computer' comes from the architecture of the RC which features a `kernel' that is a reservoir containing a large number of neurons.

Figure \ref{fig:f1}(b) schematically shows the EORC system considered which is a combination of optical (in blue) and (electronic) radio frequency (RF) (in black) systems. The EORC is a feedback system in which the signal is delayed by a time $\tau$  due to the presence of the long delay line. The delayed feedback signal is combined with the present-time input signal to modulate the MZM completing the feedback loop. The following sub-sections describe the implementation of each part of the RC as an EORC system.

\subsubsection{Input node by signal time multiplexing}
The role of the input node is to distribute the input signal $u(t)$ to the reservoir kernel. In the algorithmic reservoir computer case, one introduces a numerical input weight $\mathbf{W}^{\mathrm{in}}\in \mathbb{R}^{N_x\times N_u}$  which is a matrix of size, $N_x\times N_u$, where $N_x$ and $N_u$ are the number of neuron nodes and the number of input signals, respectively. For the translation system as an EORC, that is considered here, neuron nodes are created artificially by means of time multiplexing and masking during the information encoding process; such artificial temporal nodes have also been referred to as virtual neuron nodes \cite{appeltant2011information}. Practically this process done by the Arbitrary Waveform Generator (AWG), see Fig. \ref{fig:f1}(b). 

\begin{figure}[bp]
    \centering\includegraphics[width=0.75\textwidth]{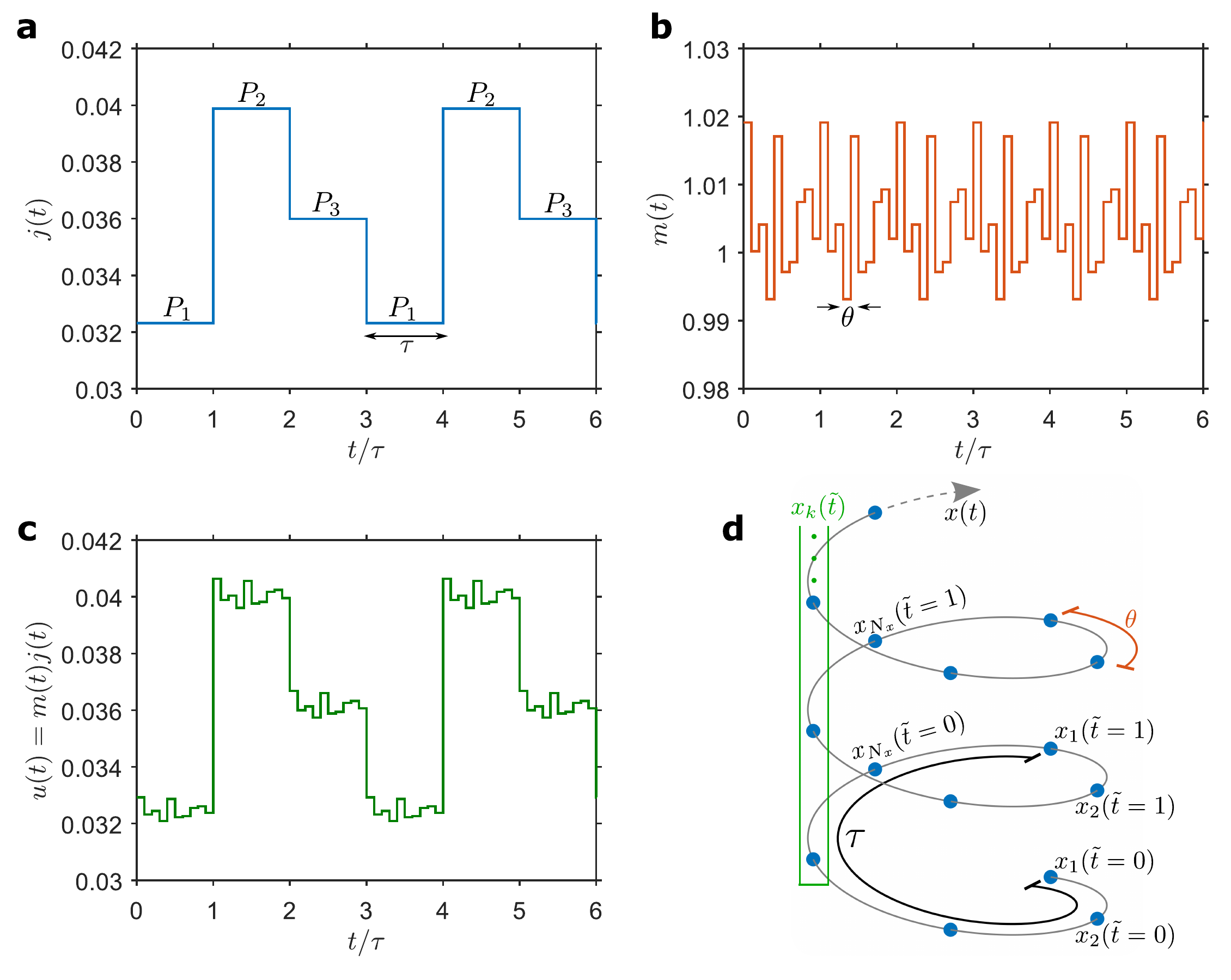}
    \caption{(a) Serialised information signal $j(t)$ which is based on the outputs of the biomimetic sensor $P_1$, $P_2$ and $P_3$; (b) Illustration of the mask signal $m(t)=m(t+\tau)$; (c) Input signal generated upon application of mask on serialised information signal; (d) Illustration of the time-demultiplexing process to generate virtual nodes.}
    \label{fig:f3}
\end{figure}

The signal time-multiplexing is achieved by performing a sample and hold operation, that is the measurement from each channel is serialised for a duration of $\tau$. For the three-channel output biomimetic sensor considered in this paper, i.e.,  $\mathbf{P}(n)=[P_1;P_2;P_3 ]^\mathrm{T}$, the serialised signal $j(t)$ is illustrated in Fig. \ref{fig:f3}(a). This time-multiplexing process allows us to encode the sensor information as a time-series signal. 

Subsequently, a periodic mask signal $m(t)=m(t+\tau)$ is generated whose values are randomly chosen from some random generator with mean of 1. This mask is piece-wise constant over the period of $\theta=\tau/N_x$. Here, the number of mask points is equal to the number of neuron nodes $N_x$,  because, as mentioned, the mask serves as the serialised input weight of the information signal with the virtual node. Figure \ref{fig:f3}(b) illustrates the mask signal for $N_x=10$. The input signal is obtained by application of mask signal to the time-multiplexed signal, i.e., $u(t)=m(t)j(t)$, see Fig. \ref{fig:f3}(c). The subsequent process applied upon input signal $u(t)$ by the `neuron' in the reservoir is described in the following. 

\subsubsection{Kernel by a single delay-line non-linear node}
The input signal $u(t)$,  {here simulated by the Arbitrary Waveform Generator}, will be combined with the delayed feedback signal $x(t-\tau)$ and transformed via a non-linear modulation process occurring in the neuron kernel. Throughout this paper, we consider a neuron kernel implementation based on the single physical node delay line with a Mach-Zehnder modulator (MZM) as the non-linear neuron activation function, see Fig.~\ref{fig:f1}(b). Implementation of a RC based on this system and its variant have been reported previously in the context of all-optical computing and communication applications \cite{appeltant2011information,argyris2018photonic}.
Initially, a similar setup to Fig.~1(a) was proposed by Ikeda \cite{Ikeda1979, Ikeda1980, sprott2003chaos} to demonstrate the existence of multiple stability state and chaos in a non-linear optical ring resonator system, mathematically, modelled by the Ikeda map given by,
\begin{align}
    v(t)=f_{\mathrm{NL}}[v(t-\tau)]
    \label{eq4}
\end{align}
where, the $f_{\mathrm{NL}}[\cdot]$ denotes a non-linear function and $\tau$ denotes the time delay parameter. Considering the specific implementation based on MZM as the non-linear function kernel, here, we have 
\begin{align}
    f_{\mathrm{NL}}[v(t)] = I_0 \cos^2\left( 
    \frac{\pi}{2V_\pi^{\mathrm{RF}}} v(t) +\Phi_0
    \right)
    \label{eq5}
\end{align}
Equation (\ref{eq5}) is the well-known non-linear transmittance of a MZM due to two-travelling wave interference process \cite{saleh2019fundamentals}.  {Parameter $I_0$ is the intensity of continuous-wave laser which is being modulated by the driving RF signal $v(t)$.}  Parameter $V_\pi^{\mathrm{RF}}$ is the half-wave voltage of the MZM i.e. the voltage which would phase shift the light by $\pi$. Parameter $\Phi_0$ is the initial phase difference between the two-travelling wave in the MZM. Commercial MZMs are usually fabricated with $\Phi_0\approx-\pi/4$, and so operates near the (almost) linear region of (\ref{eq5}); this operation is commonly referred to as the `Quad' point. For numerical simplicity, here, the half-wave voltage is taken as $V_\pi^{\mathrm{RF}}=\pi/2$.  

The original Ikeda map (\ref{eq4}) has been adapted to model a realistic experimental setup \cite{Paquot2012,Larger2013}. Specifically,  low-pass and bandpass filter effects are included to describe  the inclusion of the photodetector and RF signal amplifier, respectively \cite{Paquot2012,Larger2013}. This leads to the following delay differential equations (DDEs),
\begin{align}
    \delta \frac{\mathrm{d} v}{\mathrm{d} t} +v = r,
    \label{eq6}
\end{align}
\begin{align}
    v +
    \frac{2m}{\omega_0}\frac{\mathrm{d} v}{\mathrm{d} t} +
    \frac{1}{\omega_0^2}\frac{\mathrm{d}^2 v}{\mathrm{d} t^2} = 
    \frac{2m}{\omega_0}\frac{\mathrm{d} r}{\mathrm{d} t},
    \label{eq7}
\end{align}
where $r(t):=x(t-\tau)+u(t)=f_\mathrm{NL}[\gamma v(t-\tau)]+u(t)$. The Ikeda map with the inclusion of the low-pass filter is given by (\ref{eq6}) and the bandpass filter effect is described by (\ref{eq7}). Parameter $\delta$ in (\ref{eq6}) defines the high-frequency cut-off of the low-pass filter $f_H=1/(2\pi \delta)$. In the bandpass filter case of (\ref{eq7}), the filter parameters $m=\frac{1}{2} \sqrt{f_H/f_L}$ denotes the damping factor and $\omega_0=2\pi\sqrt{f_L f_H}$ denotes the internal pseudo-oscillator resonant frequency of the bandpass filter whose values depend on the low-frequency $f_L$ and high-frequency $f_H$ cut-off of the filter, see Appendix 5.1.  {We note that while current photodetectors can have a very high cut-off frequency, e.g., up to 20 GHz\cite{gandh}, in practice the bandwidth is further limited by the electronics setup, e.g., RF amplifier which is a bandpass filter\cite{ixblue}.} It is noted here that (\ref{eq7}) has also been formulated as an integro-differential delay equation (iDDE) as shown in \cite{Larger2013}.The impact of filtering will be discussed in detail in Section~3.1. 

The DDEs (\ref{eq6},\ref{eq7}) have been represented as a signal flow diagram in Fig. \ref{fig:f1}(c). Figure \ref{fig:f1}(c) translates the mathematical operation performed at each stage of the signal’s flow and uses the notation that will be used throughout this paper. Parameter $\gamma$ has been added to represent the effective gain of the system. Operator $h(t)$ represents the time-domain operation of the filter effect, implementing DDE (\ref{eq6}) or (\ref{eq7}). In this paper, we have implemented a time-stepping algorithm, performing the integration of (\ref{eq6}) or (\ref{eq7}) up to second-order accuracy, based on the $\mathcal{Z}$-bilinear transformation technique. This technique has previously used in conjunction with the transmission-line method to model non-linear  photonic materials described by a Duffing model \cite{Phang2017a, Phang2017b, Phang2014,phang2016theory}. Specific implementation of the filters (\ref{eq6}) and (\ref{eq7}) is provided in Appendix 5.1. Other methods \cite{sprott2003chaos, Cryer1972, bellen2013numerical} could also be used.

The activation state signal $x(t)$, as the observable, is monitored and demultiplexed for further weighting by the Read-out layer. The subsequent sub-section describes the time-demultiplexing process, and the weight applied to the activation state signal.

\subsubsection{Read-Out layer}
To allow weighting by the Read-Out layer, the activation state signal $x(t)$ is sampled and demultiplexed. Figure \ref{fig:f3}(d) schematically illustrates the time-demultiplexing process. The neuron activation state signal $x(t)$ is sampled with a sampling-time of $\theta$ and spun with a winding period of $\tau$. The demultiplexed activation signal is partitioned as $\mathbf{x}(\tilde{t})=[x_1 (\tilde{t}),x_2 (\tilde{t} ),\cdots,x_k (\tilde{t})]^\mathrm{T}$ where $k=1,2,3,\cdots,N_x$ is the index of the virtual nodes and each entries of matrix $\mathbf{x}$ are given by $x_k (\tilde{t}\tau+k\theta)=x(t)$ with $\tilde{t}=0,1,\cdots$ being the winding index. 


In the following step, the demultiplexed signal is weighted by the read-out layer, 
\begin{align}
    \mathbf{y}(\tilde{t}) = \mathbf{W}^{\mathrm{out}}[1;\mathbf{x}(\tilde{t})
    ;\mathbf{y}(\tilde{t}-1) ] ^\mathrm{T},
    \label{eq8}
\end{align}
where $ \mathbf{W}^{\mathrm{out}} \in \mathbb{R}^{N_y\times (1+N_x+N_y)}$ is the read-out weighting. Note, that in (\ref{eq8}) there is constant bias of 1 which has a similar role as in a feed-forward neural network to allow shifting of discrimination plane \cite{reed1999neural}. The read-out weight $\mathbf{W}^\mathrm{out}$ is obtained following a training session. Figure \ref{fig:f1}(c) shows the flow of signals in the normal operation in black whilst the flow of signals in the training operation is shown in red. During the training operation, given the pair of activation state signal $\mathbf{x}(\tilde{t})$ and the teacher signal $\mathbf{y}_\mathrm{teacher}(\tilde{t})$, the weight of the read-out layer is optimised. For details of the optimisation algorithm used to calculate the read-out weight via Tikhonov regularisation with cross-validation, the reader is referred to \cite{phangiet2020}.

\section{Results and discussion}
In Section 2, the sensing component of the system in the form of a biomimetic sensor with 3 output channels, and the architecture of the EORC system and its model which is used to process the sensor signal have been described. This Section starts with results and discussion of the dynamics of the single delay line feedback system (sub-sections 3.1) before describing in sub-sections 3.2 and 3.3 the application of such a system as an EORC for chemical sensing detection.  

\subsection{Dynamics of electro-optic reservoir computer under limited bandwidth operation}
A delay feedback system, including the Ikeda map \cite{sprott2003chaos}, is known to exhibits chaotic behaviour. This sub-section will discuss the influence of the filtering effects that would be present in a practical EORC system, due to the inclusion of photodetector and RF signal amplifier. For this purpose, we consider the case that there is no information signal to be processed, i.e., $u(t)=0$.

\begin{figure}[tbp]
    \centering\includegraphics[width=0.98\textwidth]{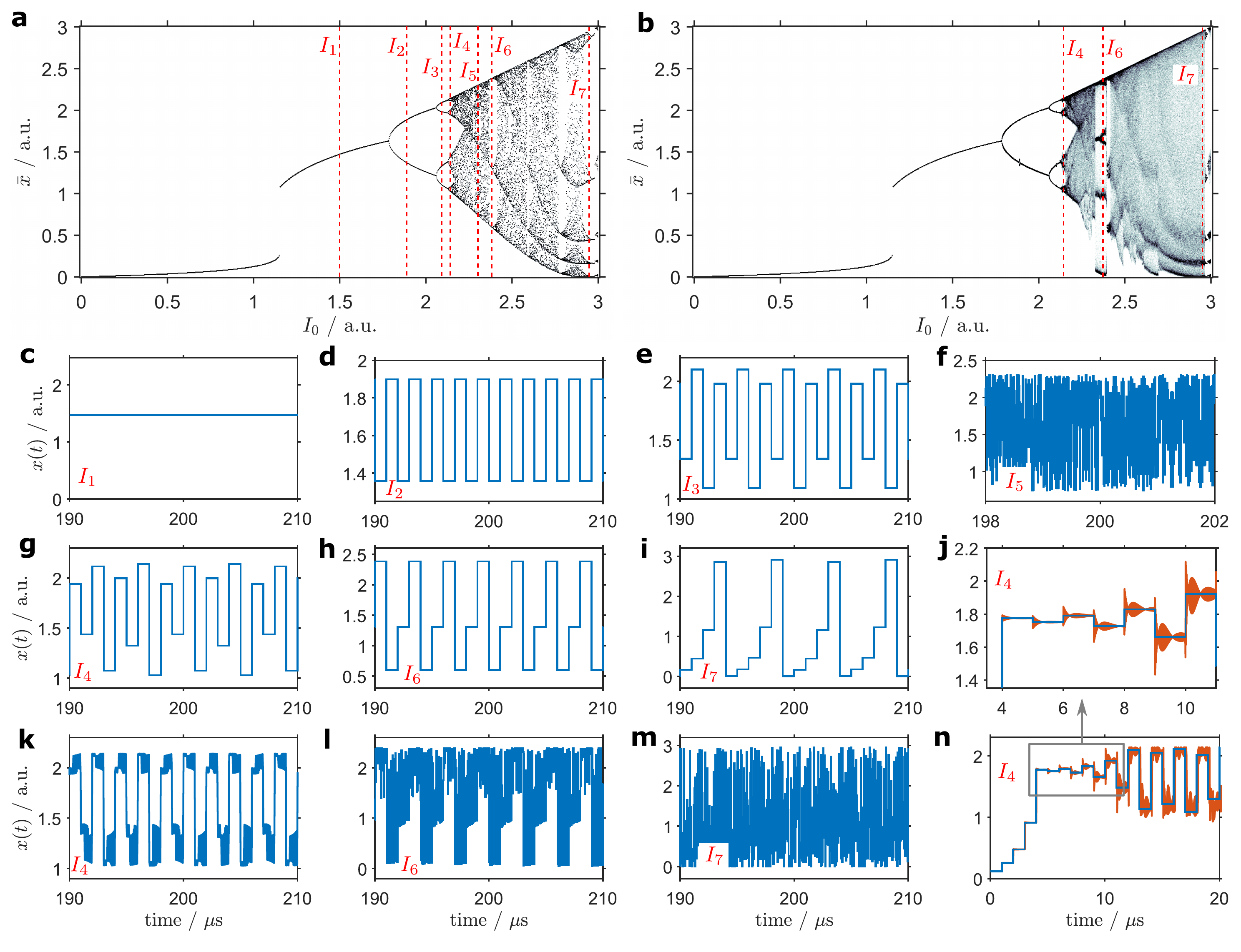}
    \caption{(a,b) Bifurcation diagram of Ikeda map (\ref{eq4}) and (\ref{eq6}), respectively. (c,d,e,f,g,h,i) The temporal signal $x(t)$ at the steady-state for different  {laser} intensity parameters of $I_1=1.5$, $I_2=1.9$, $I_3=2.1$, $I_4=2.14$, $I_5=2.3$, $I_6=2.38$ and $I_7=2.95$ for Ikeda map (\ref{eq4}). Subfigure (j) is an enlarged view of (n). (k,l,m) Temporal signal $x(t)$ at the steady-state for $I_4$, $I_6$  and $I_7$ for low-pass feedback Ikeda system (\ref{eq6}). (n) Transient state of $x(t)$ for Ikeda map (\ref{eq4}) in blue and low-pass feedback Ikeda system (\ref{eq6}) in red. For convenience, the intensity parameters considered are marked in (a,b).}
    \label{fig:f5}
\end{figure}

Figure \ref{fig:f5}(a) shows the bifurcation diagram of the original Ikeda map (\ref{eq4}). The bifurcation diagram shows all possible values for neuron activation signal $x(t)$ at the steady state as a function of laser intensity parameter $I_0$,  {also see Fig. \ref{fig:f1}(c) for notations used}. Specific parameters used for Fig. \ref{fig:f5}(a) are $\tau=1$ $\mu$s, $\gamma=1$ and $\Phi_0=-0.85\pi$ similar to that use in \cite[Fig. 3]{Larger2013}.  {Such a delay $\tau=$ 1 $\mu$s can, in practice, be achieved by using a $\approx$200 m spool of a standard SMF-28 fibre.} Signals $x(t)$ for intensities of $I_1=1.5$, $I_2=1.9$ and $I_3=2.1$ are shown in Fig. \ref{fig:f5}(c,d,e); these specific intensity points  are marked in Fig. \ref{fig:f5}(a) for reference. When operating at $I_1$, the neuron signal is single valued around 1.5, while for the case of $I_2$ and $I_3$ the signal $x(t)$ are periodic and multivalued, i.e., two and four stationary states, respectively. Figure~\ref{fig:f5}(a) further shows that while at low intensity the steady state of neuron activation signal $x(t)$ is single valued, it splits as $I_0$ increases; the splitting point is called the bifurcation point. Furthermore, Fig.~\ref{fig:f5}(a) shows that by increasing the parameter $I_0$ the steady state solution splits further and begins to behave chaotically, also see Fig.~\ref{fig:f5}(f) for the case of $I_5=2.3$ and also, noting the bistable response at $I_0=1.16$. For detail on the bifurcation and bistablilty phenomena, and the route to chaos of the Ikeda map system, see \cite{Larger2013}.

\begin{figure}[tbp]
    \centering\includegraphics[width=0.6\textwidth]{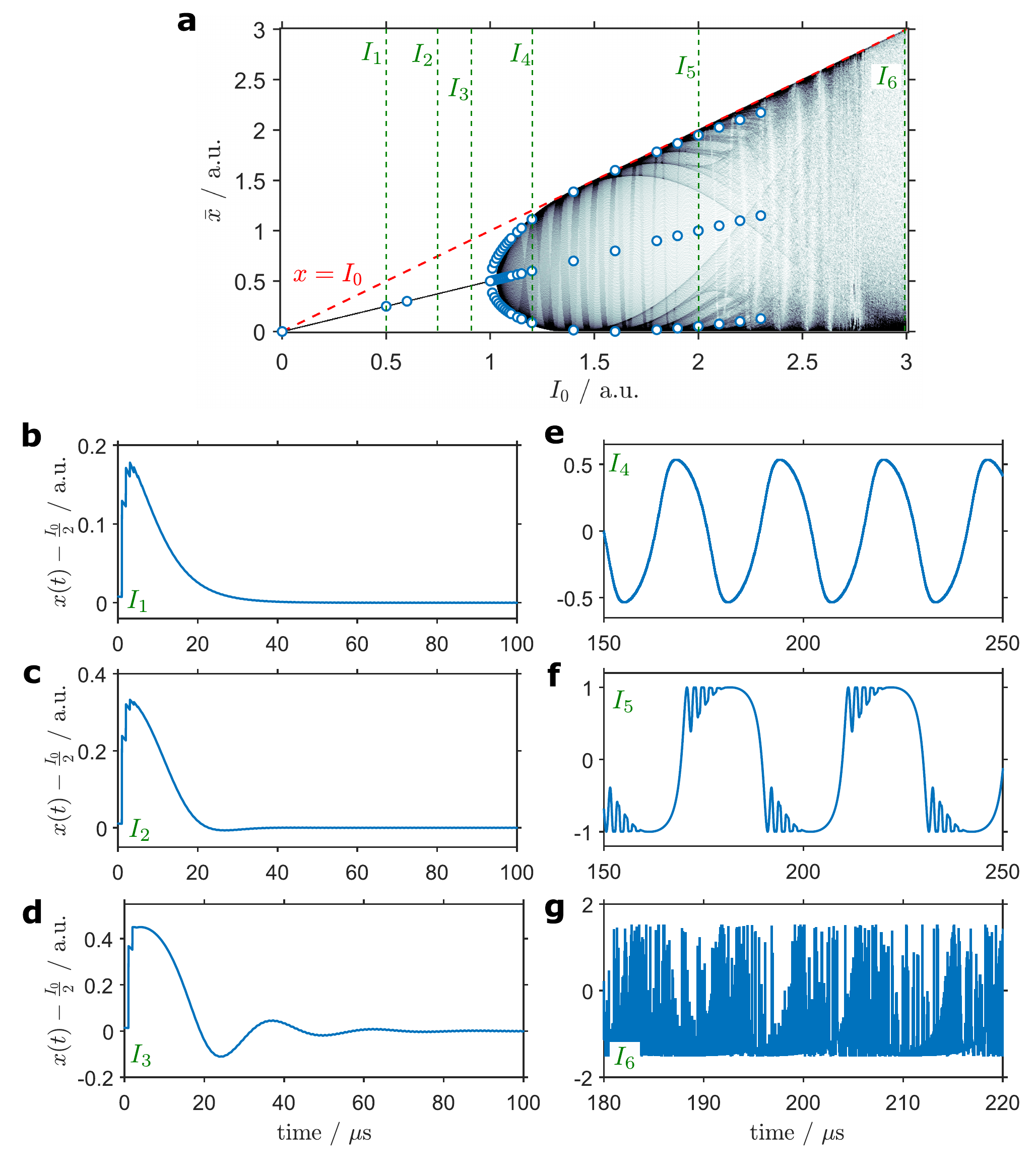}
    \caption{Bifurcation diagram of the bandpass filter feedback of eq. (7). Blue bullets points are the `fixed points' solution of (\ref{eq7}), see Appendix 5.2. The line $x=I_0$ is included for reference; (b,c,d,e,f,g) Temporal signal $x(t)-I_0/2$ at different intensity parameter of $I_1=0.5$, $I_2=0.7$, $I_3=0.9$, $I_4=1.2$, $I_5=2$ and $I_6=3$.  These intensity parameters also are marked in (a) for reference.  }
    \label{fig:f6}
\end{figure}

The bifurcation diagram for the low-pass feedback (\ref{eq6}) case is shown in Fig. \ref{fig:f5}(b). Figure~\ref{fig:f5}(b) used the same parameters as for Fig. \ref{fig:f5}(a) in addition to the high-frequency cut-off (-3dB) parameter of $f_H=10$ GHz; such a high cut-off is currently possible due to the availability of ultrafast photodetectors, e.g., InGaAs detectors \cite{gandh}. It is noted that in the absence of any filtering effect, the original Ikeda map (\ref{eq4}) operates at infinite bandwidth and as such the response of the system is instantaneous. In contrast, the DDE (\ref{eq6}), i.e., low-pass feedback system, operates with a limited bandwidth. Note that DDE (\ref{eq6}) is a causal system whose response is dependent on a past event. Visual comparison of Fig. \ref{fig:f5}(a) and Fig. \ref{fig:f5}(b) shows that the low-pass filter effect has small impact for low intensity $I_0<2.1$. The temporal signals $x(t)$ for intensity parameters of $I_1$, $I_2$ and $I_3$ are also overlapped with that of the original Ikeda map in Fig. \ref{fig:f5}(c,d,e). Deviation from the original Ikeda map in the low-pass feedback case is noted at a high-intensity whose value depends on the cut-off frequency $f_H$;  for $f_H=10$ GHz in Fig. \ref{fig:f5}(b), this occurs at $I_4\approx2.14$. Comparing Fig. \ref{fig:f5}(g) and \ref{fig:f5}(k), there is a noticeable presence of ripples at the stationary point in the low-pass feedback case but not in the infinite bandwidth case. This ripple originates from the causal response of the low-pass filter which is not instantaneous resulting a dispersion effect of the signal. This rationale is confirmed by analysing Fig. \ref{fig:f5}(n,j). Figure \ref{fig:f5}(j) is an enlargement of Fig. \ref{fig:f5}(n) which shows the transient of $x(t)$ when operated at $I_4$. It shows that, in the low-pass feedback system, signal $x(t)$ exhibits a typical dispersive ripple at signal's transition whose impact accumulates as $x(t)$ evolves over time.   

The bifurcation diagram for the case of the bandpass feedback system is shown in Fig. \ref{fig:f6}(a); the temporal signal $x(t)$ for different intensity parameters $I_i$,   $i=1,2,3,4,5,6$ are shown in Fig.~\ref{fig:f6}(b-g). The bandpass filter parameters are $f_L=10$ kHz and $f_H=10$ GHz, for the low and high-frequency cut-off, see Appendix 5.1. In addition, the MZM is operated at the Quad operational condition, i.e., $\Phi_0=-\pi/4$. At low intensity $I_0$, signal $x(t)$ reaches a steady state value of $x(t\rightarrow \infty)=I_0/2$. This is because bandpass filter removes the zero-frequency (DC) component of signal $v(t)$, as such $v(t\rightarrow \infty)=0$ leads to $x(t\rightarrow \infty)=I_0 \cos^2 (\pi/4)$. Visual inspection of Fig. \ref{fig:f6}(b – d) shows that the time required for the system to reach steady state becomes longer as the intensity increases. Then, at $I_0=1$, the neuron activation signal begins to oscillate. Figure \ref{fig:f6}(e) which depicts the signal $x(t)$, at $I_0=1.2$, oscillates at the steady state and is exhibits the skewed waveform typically found in strong non-linear media due to the presence of high-order harmonics \cite{Janyani2005}. 

\begin{figure}[bp]
    \centering\includegraphics[width=0.8\textwidth]{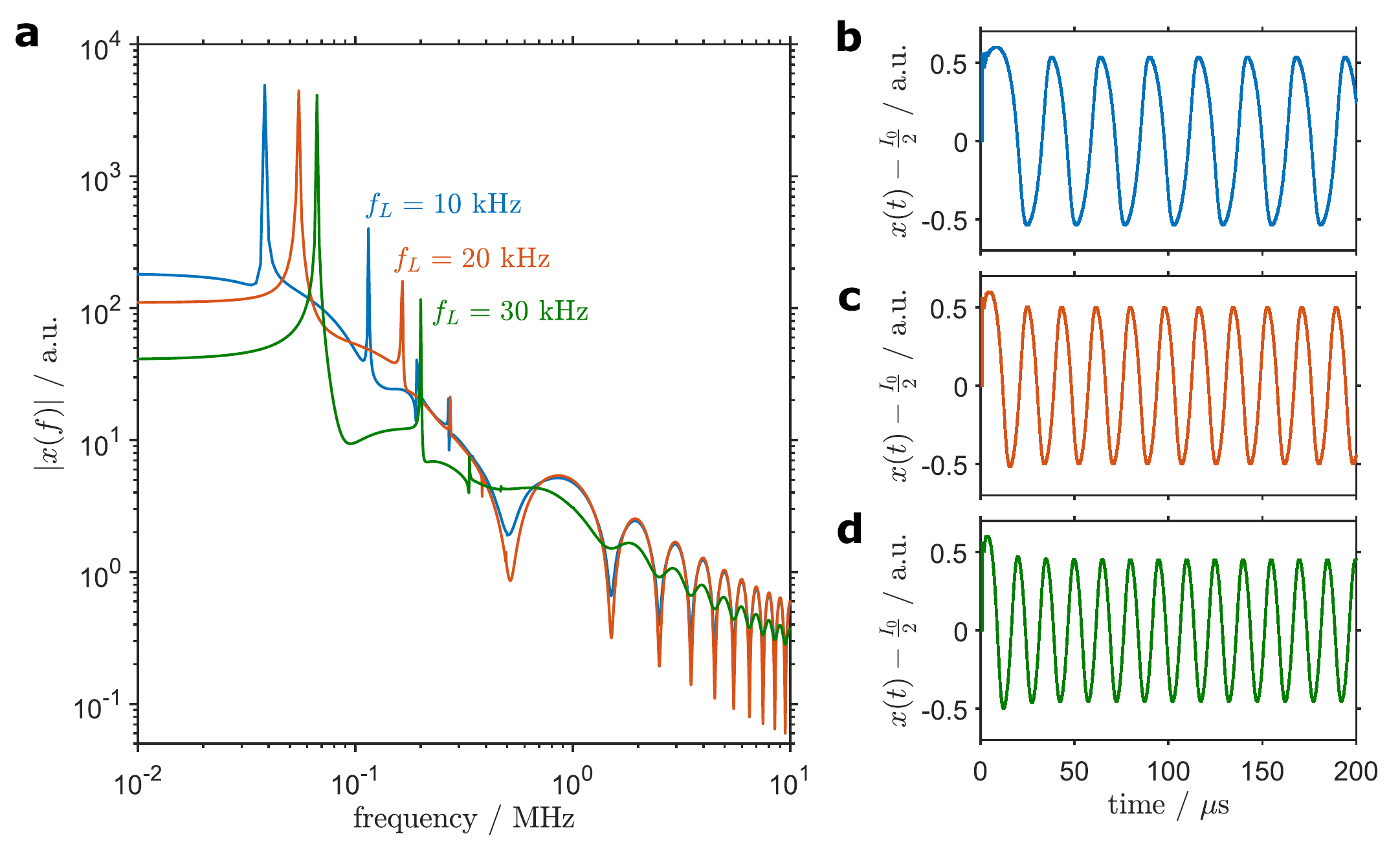}
    \caption{Impact of low-frequency cut-off $f_L$ on the dynamic of bandpass feedback system beyond the bifurcation point. (a) Spectra of $x(t)$ for different cases of $f_L=10$, 20 and 30 kHz. The temporal evolution of $x(t)$ is shown in (b,c,d), respectively. Note that in (b,c,d), $x(t)$ is shown from $t=0$ to steady state. In all cases the intensity parameter is $I_0=1.2$. }
    \label{fig:f7}
\end{figure}

Furthermore, visual inspection of Fig. \ref{fig:f6}(e) shows that the signal $x(t)$ oscillates at longer period $\approx$~25 $\mu$s than the round-trip period $\tau=1$ $\mu$s. To investigate the origin of this low-frequency oscillation, the spectrum of $x(t)$ is displayed in Fig. \ref{fig:f7}(a) for different low-frequency cut-offs of the bandpass filter, namely $f_L=10$, 20 and 30 kHz. The resulting temporal signals $x(t)$ are shown in Fig. \ref{fig:f7}(b-d), respectively. Figure \ref{fig:f7}(a) displays different characteristic behaviour in low- and high- frequency regions. The high-frequency region is characterised by the resonance of the feedback system whose fundamental frequency occurs at $1/\tau=1$ MHz and harmonics of which are produced from the nonlinear process of the MZM. Note that these harmonics are similar regardless the different $f_L$ parameters.

The characteristic behaviour in the low-frequency region is, however, significantly different for the different cases of $f_L$. Figure \ref{fig:f7}(a) shows that the fundamental frequencies of the low-frequency region are increasing, i.e., 38, 55, 67 kHz for increasing $f_L=10$, 20 and 30 kHz, confirming the origin of the low-frequency oscillation as the interplay of the $f_L$ parameter and the nonlinear feedback effect. Analytical investigations on similar bandpass feedback systems \cite{Weicker2013, weicker2012strongly} have also noted the existence of this novel interplay effect, which is attributed to the stability of the zero ‘fixed point’. As such, the fixed points of (\ref{eq7}) are shown in Fig. \ref{fig:f6}(a) as bullets points; Appendix 5.2 describes the fixed points and the bifurcation point of (\ref{eq7}) for Quad operation. For $\gamma=1$, the bifurcation point, which occurs at $I_0=1$, marks the beginning of such oscillation in $x(t)$. Further increase of intensity leads to the generation of more harmonics which manifest themselves as ripples, as shown in Fig. \ref{fig:f6}(f). Similar to the case shown in Fig. \ref{fig:f5}(a and b), further increase of intensity leads to chaotic dynamics of the system. Although not shown in Fig. \ref{fig:f6}(a), we note the existence of a metastable regime at $I_0\approx5$ after the chaotic regime.

\subsection{Sample discrimination task}
In sub-section 3.1, the impact of filter (low-pass and bandpass) on the dynamics of the EORC was discussed. In this sub-section, the EORC is trained and tasked to discriminate three chemicals, namely methanol, ethanol and isopropanol. 

To achieve such a task, the input signal has been generated following the scheme described in Section 2.2. Moreover, the teacher signal $\mathbf{y}_\mathrm{teacher}$ has been defined using one-hot-encoding, as 
\begin{align}
    \mathbf{y}_\mathrm{teacher} := [y_1;y_2;y_3]^\mathrm{T} = 
    \begin{cases}
    [1;0;0]^\mathrm{T}, \qquad \text{for methanol}\\
    [0;1;0]^\mathrm{T}, \qquad \text{for ethanol}\\
    [0;0;1]^\mathrm{T}, \qquad \text{for isopropanol}
    \end{cases}.
    \label{eq9}
\end{align}

Moreover, we only consider an EORC based on bandpass feedback system modelled via (\ref{eq7}); this is a realistic scenario as practical loss in the experimental system \cite{Larger2013,Paquot2012,appeltant2011information} demands the use of an RF amplifier. The bandpass filter parameters used throughout are $f_L=10$ kHz and $f_H=10$~GHz; such a broadband filter is achievable in practice noting the availability of ultrawide bandwidth photodetectors, e.g., \cite{gandh} and RF amplifiers, e.g., \cite{ixblue}. The delay time of the EORC used is the same as in the previous Sections $\tau=1$ $\mu$s. The impact of different numbers of virtual nodes (or number of mask points) $N_x$ is now investigated.

\begin{figure}[bp]
    \centering\includegraphics[width=0.99\textwidth]{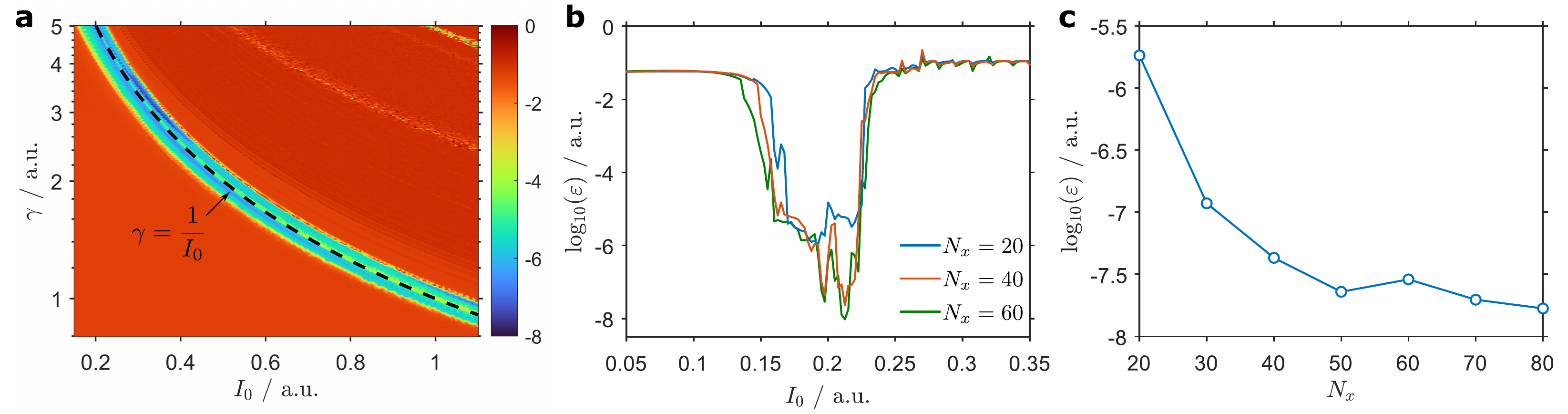}
    \caption{Impact of EORC parameters $I_0$ and $\gamma$ to the accuracy of the trichotomy task. (a) The NMSE $\varepsilon$, in logarithmic scale, is shown as the colour as function of $I_0$ and $\gamma$. The Bifurcation points are included for reference. (b) Error as function of intensity parameter $I_0$ for gain parameter $\gamma=5$; (c) Impact of number of nodes $N_x$ to the NMSE $\varepsilon$.}
    \label{fig:f8}
\end{figure}

Optimum operation of the EORC, which leads to the minimum Normalised Mean Square Error (NMSE) $\varepsilon$, occurs around the bifurcation point $I_0=1/\gamma$. It is because that around the bifurcation point the EORC is sensitive to perturbation as such is suitable for sensing/detection application. To show the relation between the intensity and gain parameter for the optimum operation of the EORC, Fig. \ref{fig:f8}(a) depicts the error $\varepsilon$ on a logarithmic scale as a function of intensity $I_0$ and effective gain $\gamma$ parameters for $N_x=30$, which is calculated by
\begin{align}
    \varepsilon = 
    \frac{\langle |\mathbf{y} -\mathbf{y}_\mathrm{teacher} |^2 \rangle}
    {\langle |\mathbf{y}_\mathrm{teacher}|^2 \rangle},
    \label{eq10}
\end{align}
where $|\cdot|$ denotes the Euclidean norm operator and $\langle \cdot \rangle$  is the assembled averaging operation over the datapoints $\tilde{t}$ and chemical samples. Figure \ref{fig:f8}(b) shows the error $\varepsilon$ as a function of $I_0$  for $N_x=20$, 40 and 60; in all cases $\gamma=5$. First, it confirms that for any $N_x$ the optimum operation of the system which leads to minimum error occurs around the bifurcation point $I_0=1/\gamma =0.2$. To investigate the impact of number of virtual nodes $N_x$, Fig. \ref{fig:f8}(c) shows error $\varepsilon$ for $I_0=0.2$ and $\gamma=5$ for different $N_x$ parameters. It confirms that one can further improved accuracy of the system by increasing the number of nodes $N_x$. Hereafter, we will use $N_x=50$ as this is shown to be sufficient to provide good accuracy operation. 

\subsection{Chemical concentration determination task}
In sub-section 3.2, the EORC has been trained and tasked to detecting pure methanol, ethanol and isopropanol. In this sub-section, the EORC is trained and tasked to determine the concentration of methanol, ethanol and isopropanol mixture. To achieve such a task, the teacher signal (\ref{eq9}) is modified to,
\begin{align}
    \mathbf{y}_\mathrm{teacher} := [y_1;y_2;y_3]^\mathrm{T} = 
    [\varrho_1;\varrho_2;\varrho_3]^\mathrm{T},
    \label{eq11}
\end{align}
where $\varrho_k$ is the vol/vol fractions as defined in (\ref{eq1}) in Section 2.1; as such $\varrho_1+\varrho_2+\varrho_2=1$.

\begin{figure}[bp]
    \centering\includegraphics[width=0.75\textwidth]{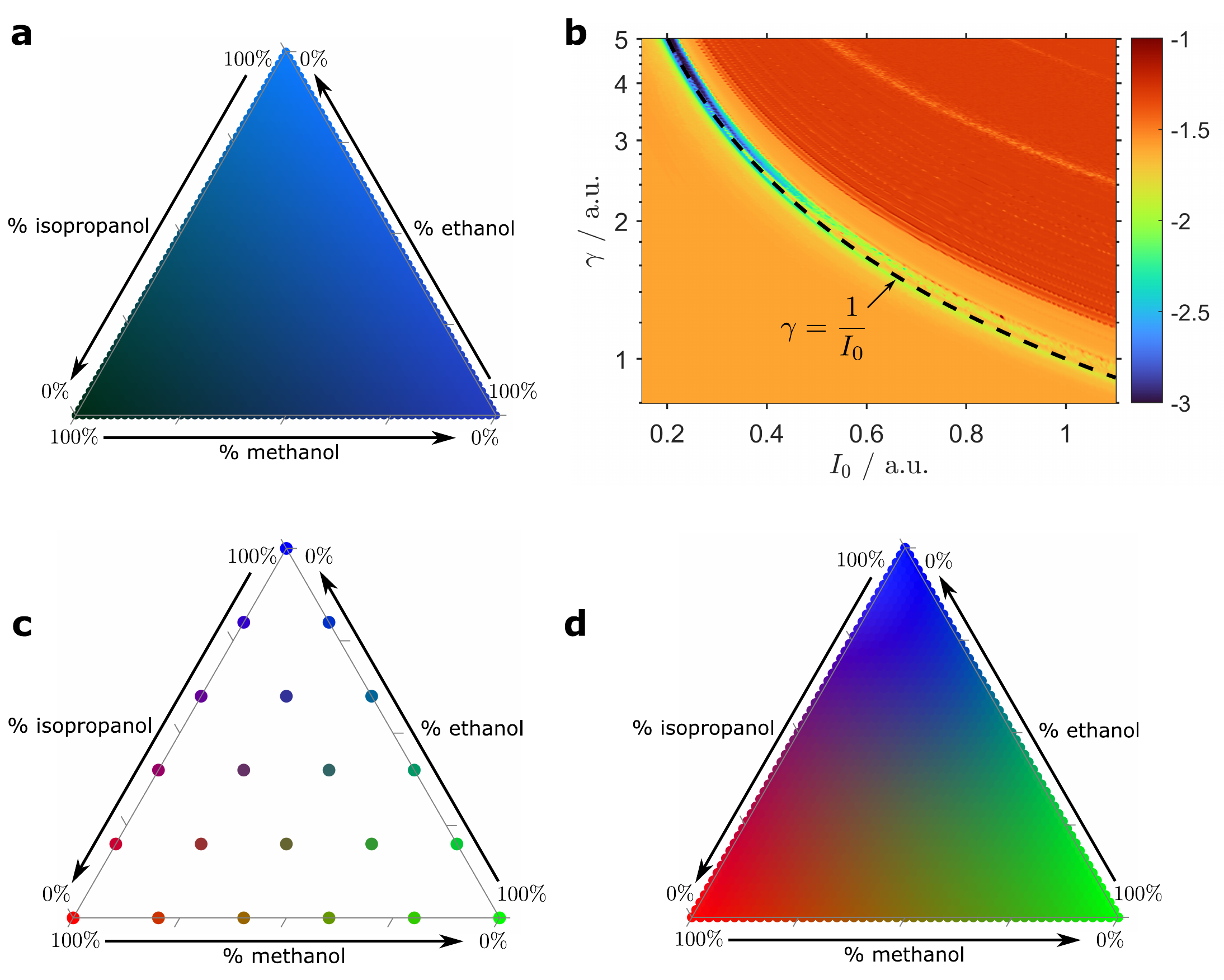}
    \caption{The EORC application for concentration determination task. (a) RGB pseudo-colouring of the output of the biomimetic sensor $\mathbf{P}(n)=[P_1;P_2:P_3]^\mathrm{T}$ for different concentration of chemical mixture shown as a ternary diagram. Note that for visual contrast $\mathbf{P}(n)=[P_1;P_2;P_3]^\mathrm{T}$ has been normalised, see sub-section 2.1. (b) Training samples of the EORC at discrete concentration combinations. (c) RGB pseudo-colouring of the steady state output signal $\mathbf{y}(t)$ for different concentration of chemical mixtures.}
    \label{fig:f9}
\end{figure}

Here, the EORC is \textbf{only} trained to recognise the concentration of few possible combinations. Figure \ref{fig:f9}(b) shows the training points used which are spaced by 20$\%$ within the ternary diagram. This specific task highlights the key capability of the artificial neural network system to perform data-interpolation and regression tasks. 

To find the optimum operation of EORC, the error $\varepsilon$ is shown as a function of intensity $I_0$ and gain $\gamma$ parameters in Fig. \ref{fig:f9}. It is similar to Fig. \ref{fig:f9}(a), with the difference being that it is based on the training at points shown in Fig. \ref{fig:f9}(b). From Fig. \ref{fig:f9}(b), the optimum operational parameters of the EORC interestingly are observed to occur at the low-intensity and high-gain $\gamma$ parameter along the bifurcation line $\gamma=1/I_0$. As such, for this task the following parameters of the EORC will be used, $I_0=0.197$, $\gamma=5$ and $N_x=50$. 

Ternary diagram, Fig. \ref{fig:f9}(d), shows the steady state of output signal $\mathbf{y}(\tilde{t})$ displayed using a RGB pseudo-colouring scheme. It is important to note that no post-processing was applied to generate Fig. \ref{fig:f9}(d).  By visual inspection, one can observe three distinct regions in Fig. \ref{fig:f9}(d) clearly indicating the presence of three underlying chemical specimens. Furthermore, one can see the change of shades of the colour-encoding as the concentration of the underlying chemicals are varied. In contrast with Fig. \ref{fig:f9}(a), there is almost no colour contrast for the case of small methanol concentration, indicating that discriminating ethanol and isopropanol mixture is a challenging task. This is perhaps understandable since, as, as seen in Fig. \ref{fig:f2}(b), the absorption spectra $\alpha$ of pure ethanol and isopropanol are similar.

\section{Conclusions}
This paper proposed a novel fully bio-inspired apparatus to monitor and determine the concentration of chemical mixture. It implements a human eyes-like sensor and neuromorphic optical system based on electro-optical implementation of reservoir computer (EORC) with the single physical node delay line and Mach-Zehnder nonlinear element for information processing. Here, a delay differential equation including a realistic bandpass filtering effect was used to demonstrate this sensing application. Two kinds of sensing application were  {simulated} successfully, namely, to discriminate known (trained) samples of pure chemicals and to determine the concentration of mixture chemical. The latter application highlights the true capabilities of EORC as member of artificial neural network family to perform interpolation and regression task.  {Experimental demonstration is currently in progress and soon will be reported.} Time-stepping algorithms based on $\mathcal{Z}$-bilinear transform technique have been developed to numerically solved the governing equations. Further optimisation that was not explored here is the optimisation of the biomimetic sensor, for example the number of channels and the wavelength of the filter, we believe such an optimisation will further increase the accuracy of the whole system.   

\section{Appendices}
\subsection{Digital filter model of (\ref{eq6} and \ref{eq7}) by $\mathcal{Z}$-bilinear transform technique}

\textbf{Low-pass digital filter}: Performing Laplace transform on the DDE (\ref{eq6}), 
\begin{align}
    \mathcal{L}[h(t)] = \frac{v(s)}{r(s)} = 
    H(s)= \frac{1}{1+\delta s}.
    \label{eq12}
\end{align}
Performing $\mathcal{Z}$-bilinear transformation $s\leftarrow \frac{2}{\Delta T} \left(\frac{1-z^{-1}}{1+z^{-1}}\right)$ on (\ref{eq12}), it can be shown
\begin{align}
    v(z)=K_1 r(z)+z^{-1} \{K_1 K_2 v(z)+K_1 r(z)\},
    \label{eq13}
\end{align}
where 
\begin{align}
    K_1=\left(\frac{2\delta}{\Delta T}+1\right)^{-1}; \qquad K_2=\frac{2\delta}{\Delta T}-1,
\end{align}
and $\Delta T$ denote the numerical accuracy of the model, throughout this paper $\Delta T=\tau/N_x$, with $N_x$ denote the number of virtual nodes.

\textbf{Band-pass digital filter}: Performing Laplace transform on the DDE (\ref{eq7}), 
\begin{align}
    \mathcal{L}[h(t)] = \frac{v(s)}{r(s)} = 
    H(s)= \frac{\frac{2m}{\omega_0} s}{1+\frac{2m}{\omega_0} s + \frac{1}{\omega_0^2}s^2}.
    \label{eq15}
\end{align}
Performing $\mathcal{Z}$-bilinear transformation $s\leftarrow \frac{2}{\Delta T} \left(\frac{1-z^{-1}}{1+z^{-1}}\right)$ on (\ref{eq15}), it can be shown
\begin{align}
    v(z)=K_1 K_2 r(z) + 
         z^{-1} \left\{ K_2 K_3 v(z) + z^{-1} \{ K_2K_4 v(z) - K_1K_2 r(z) \}\right\},
         \label{eq16}
\end{align}
where 
\begin{align}
\begin{split}
    &K_1 = \frac{4m}{\omega_0 \Delta T} ; \quad 
    K_2 = \left(1+ \frac{4m}{\omega_0 \Delta T} + \frac{4}{\omega_0^2 \Delta T^2} \right)^{-1}; \\
    &K_3= -2 + \frac{8}{\omega_0^2 \Delta T^2} ; \quad 
    K_4 = -1+ \frac{4m}{\omega_0 \Delta T} - \frac{4}{\omega_0^2 \Delta T^2},
\end{split}
\end{align}
and $\Delta T$ denote the numerical accuracy of the model, throughout this paper $\Delta T=\tau/N_x$, with $N_x$ denote the number of virtual nodes.

\begin{figure}[!ht]
    \centering\includegraphics[width=0.9\textwidth]{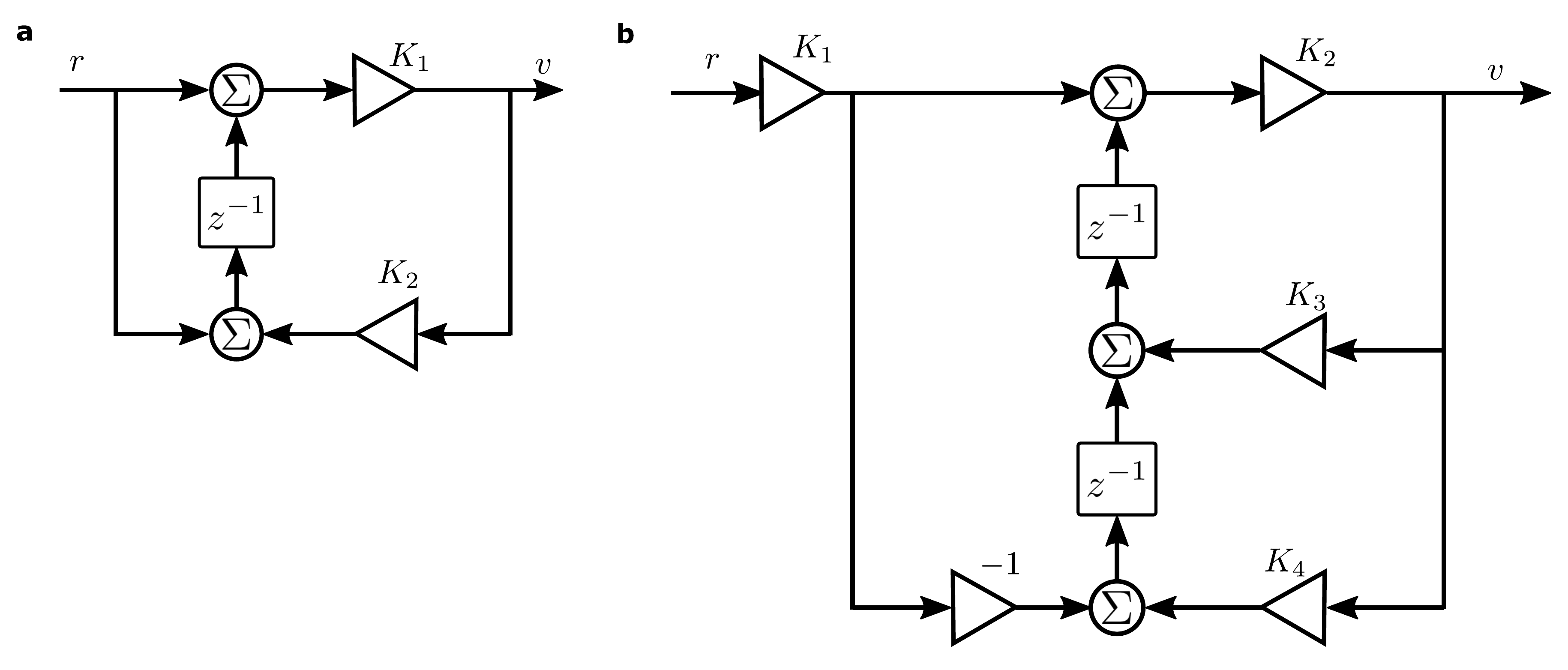}
    \caption{Digital filter design for the low-pass filter (a) and the bandpass (b) filter used to model DDEs (\ref{eq6}) and (\ref{eq7}), respectively. }
    \label{fig:f10}
\end{figure}

Depending on the value of parameter $m$, (\ref{eq15}) can model either a narrow or broad band-pass filter \cite{Larger2013}. With $m<1$, (\ref{eq15}) has complex eigenvalues exhibiting a high-Q factor oscillator, as such operates as a narrow band-pass filter centred at $\omega_0$. For $m>1$, the eigenvalues of (\ref{eq15}) are real and negative valued, such system is a strongly damped oscillator. The transfer function (\ref{eq15}) with $m>1$ exhibits a broad band-pass filter effect with low-frequency cut-off (-3 dB), $f_L=\omega_0/(4\pi m)$ and the high-frequency cut-off (-3 dB) $f_H=(m \omega_0)/\pi$, from which filter parameters $m=\frac{1}{2} \sqrt{f_H/f_L}$ and $\omega_0=2\pi\sqrt{f_L f_H}$.  The digital filter circuit of (\ref{eq13}) for the low-pass and (\ref{eq16}) for the bandpass filter is depicted in Fig. \ref{fig:f10}(a and b), respectively.    

\subsection{Stationary and bifurcation point of (\ref{eq7}) at the Quad operational condition}
Taking an adiabatic limit at the steady state reduces (\ref{eq7}) to
\begin{align}
    x(t+\tau)= v(t+\tau)+I_0  \cos^2 \Phi_0 = I_0  \cos^2 \left(\gamma v(t)+ \Phi_0 \right).
    \label{eq18}
\end{align}
For specific `Quad' operational point at positive slope of (\ref{eq5}), one can calculate the `fixed point' $\bar{v}(I_0,\Phi_0=-\pi/4)$ by solving,
\begin{align}
    \bar{v}=I_0 \cos^2\left( \gamma \bar{v} -\frac{\pi}{4}\right) - \frac{I_0}{2}. 
\end{align}
In the presence of band-pass filter, the steady state of signal $v(t)$ will has a mean value of zeros. Thus, one can expand the non-linear term in its series form, from which the stationary point of (\ref{eq18}) can be approximated by,  
\begin{align}
    \bar{v}=I_0\left(\frac{1}{2} +\gamma \bar{v} -\frac{2}{3} \gamma^3 \bar{v}^3\right)
    -\frac{I_0}{2} \rightarrow \bar{v} =\left\{0;\pm \frac{3}{2\gamma} \sqrt{\gamma - \frac{1}{I_0}} \right\}.
    \label{eq20}
\end{align}
Substituting (\ref{eq20}) back to (\ref{eq18}), 
\begin{align}
    \bar{x} =\left\{\frac{I_0}{2};\frac{I_0}{2} \pm \frac{3}{2\gamma} \sqrt{\gamma - \frac{1}{I_0}} \right\}.
\end{align}
Specifically, the bifurcation point occurs at, 
\begin{align}
    \gamma = \frac{1}{I_0}.
\end{align}
The analytical solution for the general case of (7) can be found in \cite{Weicker2013}.

\begin{backmatter}
    \bmsection{Funding}
    D.F. acknowledges the support of the Engineering and Physical Sciences Research Council (EP/P013708/1, EP/T010762/1). 
    
    \bmsection{Disclosures}
    The authors declare no conflicts of interest.
    
    \bmsection{Data availability} Data underlying the results presented in this paper are not publicly available at this time but may be obtained from the authors upon reasonable request.
\end{backmatter}


\bibliography{myrefs}

\end{document}